\documentclass[12pt,a4paper]{article}

\usepackage{amsfonts}
\usepackage{amssymb}
\usepackage[dvips]{graphicx}
\usepackage[usenames]{color}

\title{\bf Forced Nonlinear Resonance in a System of Coupled Oscillators}

\author{Sergei Glebov \thanks{Ufa State Petroleum Technical University
                              ({\tt glebskie@gmail.com})}
        \and
        Oleg Kiselev \thanks{Institute of Mathematics USC RAS
                             ({\tt ok@ufanet.ru})}
        \and
        Nikolai Tarkhanov \thanks{Institute of Mathematics, Potsdam University
                                  ({\tt tarkhanov@math.uni-potsdam.de})}
       }

\date{December 7, 2010}

\begin{document}

\maketitle

\begin{abstract}
We consider a resonantly perturbed system of coupled nonlinear oscillators with small dissipation and outer periodic perturbation.
We show that for the large time $t \sim \varepsilon^{-2}$ one component of the system is described in the main by the inhomogeneous Mathieu equation while the other component represents pulsation of large amplitude.
A Hamiltonian system is obtained which describes the behaviour of the envelope
in the main.
The analytic results agree to numerical simulations.
\end{abstract}

\section{Preliminaries}
\label{s.preliminaries}
\setcounter{equation}{0}

The paper deals with a system of coupled oscillators.
This system has a large number of applications.
In the framework of classical theory it can be used e.g. for describing oscillations of electrons and atoms under forced combinative scattering,
   see \cite{Blombergen}-\cite{Akhmanov}.

For instance consider isotropic medium and denote
   by $u$ the normal coordinate of oscillations of electrons in atoms and
   by $v$ the normal coordinate of oscillations of atoms in molecules.
The frequency of oscillations $u$ corresponds to the ultraviolet scale range
while the frequency of oscillations $v$ lies in the ultrared scale range.
To a linear approximation the oscillations of electrons and atoms occur
independently of each other.

In the general case on taking third order terms into account the expression for the potential energy has the form
$$
   U (u,v)
 = \frac{1}{2} \alpha u^2
 + \frac{1}{2} \beta v^2
 + \beta_1 v^3
 + \beta_2 u v^2
 + \beta_3 u^3
 + \frac{1}{2} \gamma u^2 v,
$$
where
   $\alpha$ and $\beta$ are constraint elasticity coefficients in a molecule,
   the coefficient $\beta_1$ determines the nonlinearity of quasielasticity of
   oscillations $v$,
   $\beta_2$ determines the parametric stimulation of oscillations $v$ by means
   of electron oscillations,
   $\beta_3$ determines the nonlinearity of electron oscillations which are
   responsible for the generation of optical harmonics.
The coefficient $\gamma$ corresponds to the nonlinear interaction of electrons and nucleus and it determines the process of combinative scattering.
Here we will consider the process of forced combinative scattering which is related to the presence of antiharmonic terms in the expression for the potential energy $U$.
We assume in the sequel that the constants $\alpha$,
                                           $\beta$,
                                           $\beta_1$,
                                           $\beta_2$,
                                           $\beta_3$
are essentially less than $\gamma$.

Under the assumptions that both $u$ and $v$ are small we get the system
\begin{eqnarray}
\label{sysAkh}
   u'' + \nu_1 u' + \omega^2 u
 & = &
   \frac{e}{m} E - \frac{\gamma}{m} uv,
\nonumber
\\
   v'' + \nu_2 v' + \mathit{\Omega}^2 v
 & = &
   - \frac{\gamma}{2M} u^2,
\nonumber
\end{eqnarray}
see \cite{Khokhlov}.
Here
   $m$ and $M$ are reduced masses of electronic and atomic oscillators,
   $\omega = \sqrt{\alpha/m}$ and
   $\mathit{\Omega} = \sqrt{\beta/M}$ are eigen oscillation frequences,
   $\gamma$ is the constant that describes the nonlinear interaction of
   electrons and nucleus,
   $e$ is the electron charge,
   $E$ stands for the electric field of light waves, and
   $\nu_1$, $\nu_2$ are dissipation coefficients.
On assuming the atomic dissipation to be essentially less that the electronic
dissipation, we will actually neglect $\nu_2$ in the sequel.

Change the dependent variables in system (\ref{sysAkh}) by
   $x = - u$ and
   $y = - v$.
On taking the radiation damping in an electronic oscillator into account,
   see \cite{Khokhlov},
       \cite{Sivukhin},
we arrive at the system
\begin{eqnarray}
\label{caupOscil}
   x''_{tt} + \nu x'_t + \omega^2 x
 & = &
   \epsilon\, x y + A\, \cos \Big( \frac{\mathit{\Omega} t}{2} \Big),
\\
   y''_{tt} + \mathit{\Omega}^2 y
 & = &
   \epsilon \delta\,  x^2,
\nonumber
\end{eqnarray}
where
   $\epsilon = \gamma/m$ is a small positive parameter,
   $\delta = m/(2M)$,
   the quantity $(e/m) E = - A \cos \left( \mathit{\Omega} t/2 \right)$ is
   periodic in $t$,
and $\omega$,
    $\mathit{\Omega}$,
    $A$,
    $\nu$
are positive constants.

Other examples of systems of coupled oscillators concern the description of
propagation of surface and interior gravitational waves, see \cite{McGoldrich},
                                                             \cite{Ball}.

The simplest analogue of system (\ref{caupOscil}) is a nonlinear oscillator with resonance pumping.
The behaviour of such an oscillator is characterised at large times by beginning
of permanent periodic nonlinear oscillations of the envelope whose amplitude is large when compared with the pumping.
This process is described by the primary resonance equation, see for instance
   \cite{Bogolyubov},
   \cite{Naife}.

The system (\ref{caupOscil}) belongs to the other class of resonantly perturbed problems.
The pumping fails here to occur directly, i.e. by including the perturbation of eigen frequency to the right-hand side of the equation, but rather by means of nonlinear interaction of oscillators.
In this situation the Mathieu equation proves to be of crucial role.

In the paper an analogue of the nonlinear resonance equation for the envelope of resonant component is derived in the case of special nonlinear coupling in system (\ref{caupOscil}).

\begin{figure}[!t]
\begin{center}
\includegraphics[width=10cm,height=6cm]{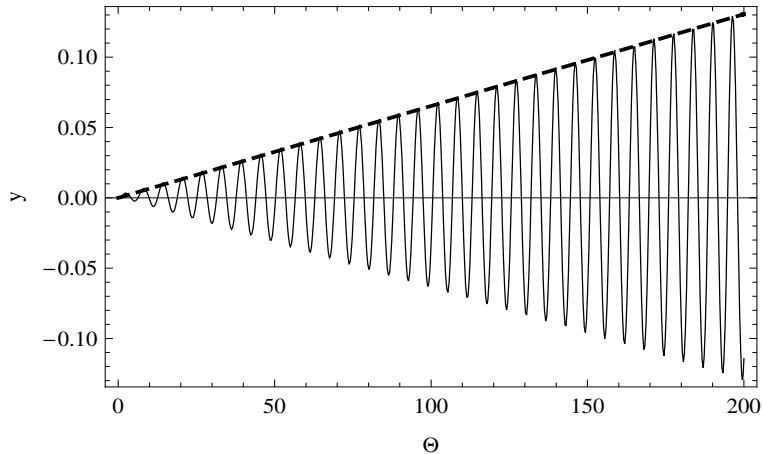}
\end{center}
\caption{Linear growth of the $y\,$-component of solution of (\ref{caupOscil}) in the initial stage. The solution amplitude is well approximated by the straight line $y = 4 \varepsilon A^2 t / (4 \omega^2 - \mathit{\Omega}^2)^2$.}
\label{fig1a}
\end{figure}

In the initial stage the solution amplitude growths linearly, see Fig. \ref{fig1a}.
This growth corresponds to the linear resonance in the second equation of the system.
The envelope is well approximated by the straight line
   $y = 4 \varepsilon A^2 t / (4 \omega^2 - \mathit{\Omega}^2)^2$.
For large values of $y$ the nonlinear effects of interaction between oscillators become essential.

\begin{figure}[!t]
\begin{center}
\includegraphics[width=10cm,height=6cm]{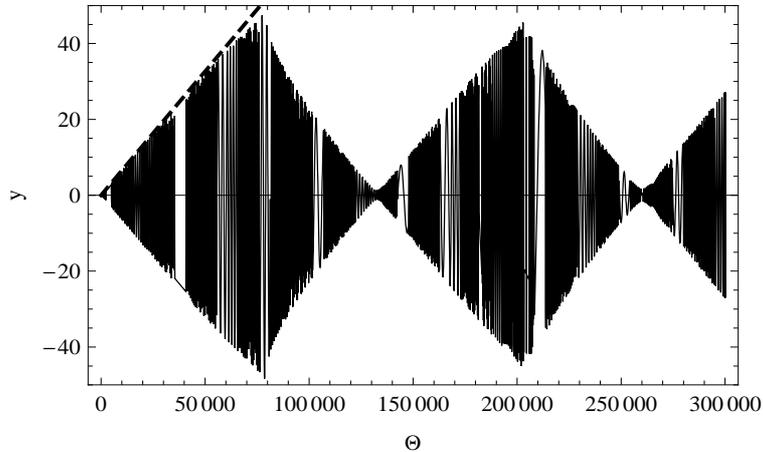}
\end{center}
\caption{Numerical simulation of the behaviour of system (\ref{caupOscil}) with  $\epsilon = 0.2, 
 \mathit{\Omega} = 1, \omega = 3, A = 1$ and $\nu = 0.1$. One can see pulsations of the $y\,$-component of solution of (\ref{caupOscil}). In the initial stage the solution amplitude is well approximated by the straight line $y = 4 \varepsilon A^2 t / (4 \omega^2 - \mathit{\Omega}^2)^2$. For large values of $y$ the nonlinear effects of interaction between oscillators become essential.}
\label{fig1}
\end{figure}

In Fig. \ref{fig1} the results of numerical simulation of the behaviour of system (\ref{caupOscil}) of weakly coupled oscillators are shown under zero initial data.
The parameter values are $\epsilon = 0.2$,
                         $\mathit{\Omega} = 1$,
                         $\omega = 3$,
                         $A = 1$ and
                         $\nu = 0.1$
Numerical simulations are realized by the Adams method.

The main result of the paper consists in describing dynamics of the envelope of oscillations for large times.
It turns actually out that the envelope oscillates as well.
We show formulas for the amplitude and period of these oscillations.

The paper consists of 5 sections.
In Section \ref{Result} we formulate the results obtained.
In Section \ref{Construction} an asymptotic solution to the initial system (\ref{caupOscil}) is constructed.
In Section \ref{lambdaBig} we study in detail the physically interesting case where the frequency quotient $\omega/\mathit{\Omega} \gg 1$ is large enough.
In this case explicit formulas for the amplitude and period of oscillations are given.
In the last Section \ref{s.conclusion} we shortly comment on the results.

\section{Results}
\label{Result}
\setcounter{equation}{0}

The asymptotic solution of system (\ref{caupOscil}) for the large time
    $t = O (\varepsilon^{-2})$
has the form
\begin{eqnarray*}
   x (t,\varepsilon) & \sim & x_0 (t,\varepsilon^2 t),
\\
   y (t,\varepsilon) & \sim & \varepsilon^{-1}\, Y (t,\varepsilon^2 t)
\end{eqnarray*}
for $\varepsilon \to 0$, where
$
   \displaystyle
   \varepsilon = \frac{\epsilon}{\mathit{\Omega}^2}
$
and
$$
   Y (t,\varepsilon^2 t)
 = 2\, \Re\, k (\varepsilon^2 t) \exp (\imath \mathit{\Omega} t),
$$
$\Re z$ standing for the real part of a complex number $z$.
The function $x_0$ satisfies the inhomogeneous Mathieu equation
\begin{equation}
\label{nonHomMathieu}
   \partial^2_{ss} x_0
 + 4 \frac{\nu}{\mathit{\Omega}}\, \partial_s x_0
 + 4 (\lambda^2 - |k|\, \cos 2s) x_0
 = 4 \frac{A}{\mathit{\Omega}^2}\, \cos \Big( s - \frac{1}{2} \arg k \Big),
\end{equation}
where
$$
   s = \frac{\mathit{\Omega} t + \arg k}{2},\ \ \ \ \
   \lambda = \frac{\omega}{\mathit{\Omega}}.
$$

Write $k = k_1 + \imath k_2$, where $k_1 = \Re k$ and
                                    $k_2 = \Im k$
are the real and imaginary parts of the function $k$, respectively.
They are solutions of the averaged system
\begin{eqnarray*}
   \partial_{\tau} k_1
 & = &
 -\, \lim_{T \to \infty}
     \frac{\mathit{\Omega} \delta}{T}
     \int_0^{T} x_0^2 (\vartheta,\tau) \sin \vartheta\, d \vartheta,
\\
   \partial_{\tau} k_2
 & = &
   \,\,\,\,\,
   \lim_{T \to \infty}
   \frac{\mathit{\Omega} \delta}{T}
   \int_0^{T}
   x_0^2 (\vartheta,\tau) \cos \vartheta\, d \vartheta,
\end{eqnarray*}
where $\tau = \varepsilon^2 t$.

In the case $\lambda = \omega/\mathit{\Omega} \gg 1$ which is important for applications the last system simplifies and reduces to the Hamiltonian system with Hamiltonian of form
$$
   H(K_1,K_2) = \frac{A^2}{2 \pi \mathit{\Omega}^3}\,
       \frac{|K|^2 - K_1}
            {K_1 (1 - |K|^2) - (|K|^2 - K_1) \sqrt{1-|K|^2}}
$$
and
$
   \displaystyle
   K_{1} = \frac{k_{1}}{\lambda^2},
$
$
   \displaystyle
   K_{2} = \frac{k_{2}}{\lambda^2}.
$

The maximal value of the amplitude $y$ for $H = H_0$ and
                                       for $t = O (\varepsilon^{-2})$
with $\varepsilon \to 0$ is evaluated by
$$
   \max |y|
 \sim \frac{\lambda^2}{\varepsilon}\, \max \{ |r_-|, |r_+| \},
$$
where $\lambda \gg 1$ and
$$
   r_\pm
 = \frac{(A^2 - 2 \pi \mathit{\Omega}^3 H_0)
     \pm \sqrt{- A^4 + 4 A^2 \pi \mathit{\Omega}^3 H_0
                     + 4 \pi^2 \mathit{\Omega}^6 H_0^2}}
        {4 \pi \mathit{\Omega}^3 H_0}.
$$
The period is in turn determined by the formula
$$
   T (H_0)
 \sim \frac{\lambda^8}{\varepsilon^{2}}\,
      \int_{c} \frac{d K_{1}}{\displaystyle \frac{\partial H}{\partial K_{2}}},
$$
the integration being over the cycle $c$ in the complex plane of the variable
$K = K_1 + \imath K_2$ given by $H(K_1,K_2)=H_0$.

\section{Formal constructions for $\varepsilon \ll 1$}
\label{Construction}
\setcounter{equation}{0}

In this section we write down the problems for determining the coefficients of
asymptotic solutions to system (\ref{caupOscil}).
For convenience, we introduce a new independent variable
   $\theta = \mathit{\Omega} t$.
On designating
$$
\begin{array}{cccccccc}
   \displaystyle
   \lambda = \frac{\omega}{\mathit{\Omega}},
 & &
   \displaystyle
   \varepsilon = \frac{\epsilon}{\mathit{\Omega}^2},
 & &
   \displaystyle
   f = \frac{A}{\mathit{\Omega}^2},
 & &
   \displaystyle
   \mu = \frac{\nu}{\mathit{\Omega}},
\end{array}
$$
we get
\begin{equation}
\label{caupOscil1}
\begin{array}{rcl}
   x''_{\theta \theta} + \mu x'_{\theta} + \lambda^2 x
 & =
 & \varepsilon\, xy + f\, \cos \frac{\theta}{2},
\\
   y''_{\theta \theta} + y
 & =
 & \varepsilon \delta\, x^2.
\end{array}
\end{equation}

The solution of this system is constructed by the two scales method
   (see \cite{Naife})
in the form
\begin{equation}
\label{secondEx}
\begin{array}{rcl}
   x (\theta,\tau,\varepsilon)
 & =
 & x_0 (\theta,\tau),
\\
   y (\theta,\tau,\varepsilon)
 & =
 & \varepsilon^{-1}\, Y (\theta,\tau) + \varepsilon\, y_1 (\theta,\tau),
\end{array}
\end{equation}
where $\tau = \varepsilon^2 \theta$ is a slow variable.

Let us substitute (\ref{secondEx}) into system (\ref{caupOscil1}) and group the coefficients of the same powers of the small parameter $\varepsilon$.
The equation for $Y$ takes obviously the form
$$
   \partial^2_{\theta \theta} Y + Y = 0.
$$
The general solution of this equation is of the form
$$
   Y (\theta,\tau)
 = k (\tau) \exp (\imath \theta)
 + \overline{k (\tau)} \exp (- \imath \theta)
$$
with arbitrary function $k (\tau)$ to be chosen later on.

The main term $x_0$ is determined from the equation
$$
   \partial^2_{\theta \theta} x_{0}
 + \mu \partial_{\theta} x_0
 + (\lambda^2 - Y)\, x_0
 = f\, \cos \frac{\theta}{2},
$$
which can be rewritten in the form
\begin{equation}
\label{nonHomMathieum}
   \partial^2_{ss} x_0
 + 4 \mu\, \partial_s x_0
 + (q - 2 r\, \cos 2s) x_0
 = 4 f\, \cos \Big( s - \frac{1}{2} a \Big)
\end{equation}
with
$$
\begin{array}{rclcrcl}
   s
 & =
 & \displaystyle \frac{\theta + a}{2},
 &
 & a
 & =
 & \arg k,
\\
   q
 & =
 & 4 \lambda^2,
 &
 & r
 & =
 & 2\, |k|,
\end{array}
$$
cf. (\ref{nonHomMathieu}).

Let $\varphi_1$,
    $\varphi_2$
be a fundamental system of solutions of the homogeneous equation corresponding to (\ref{nonHomMathieum}).
Then any solution of (\ref{nonHomMathieum}) can be represented in the form
$$
   x_0
 = c_{1} \varphi_1 + {c}_{2} \varphi_2
 + \int_{s_0}^{s}
   \frac{\varphi_1 (s) \varphi_2 (s') - \varphi_2 (s) \varphi_1 (s')}
        {\varphi_1 (s') \varphi_2' (s') - \varphi_2' (s') \varphi_1 (s')}\,
   4 f\, \cos \Big( s' - \frac{1}{2} a \Big)
   ds',
$$
the denominator being the Wronsky determinant of the linearly independent system
   $\{ \varphi_1, \varphi_2 \}$.

The change of dependent variables
   $x_0 = u\, \exp (- 2 \mu s)$
reduces the homogeneous equation corresponding to (\ref{nonHomMathieum})
\begin{equation}
\label{HomMathieu}
   \partial^2_{ss} x_0 + 4 \mu\, \partial_s x_0 + (q - 2 r\, \cos 2s) x_0
 = 0
\end{equation}
to the Mathieu equation with slowly varying coefficient $r$.
More precisely, we get
\begin{equation}
   u'' + (Q - 2R\, \cos 2s) u = 0, \label{Maa}
\end{equation}
where $Q = q - \mu^2$ and
      $R = R (\varepsilon^2 s)$.

It is known that the general solution of the Mathieu equation changes within a period as
$
u_{1,2} (s + 2 \pi) = \exp (2 \pi \lambda_{1,2}) u_{1,2} (s),
$
where $\lambda_{1,2} (q,r)$ are characteristic indices of the (homogeneous) Mathieu equation, see \cite{Floke}.
For $q$ and $r$ such that both the indices are purely imaginary, the solution
of the Mathieu equation is bounded.
If either of the characteristic indices is real and positive then the solution
has exponential growth.

\begin{figure}[!t]
\begin{center}
\includegraphics[width=10cm,height=6cm]{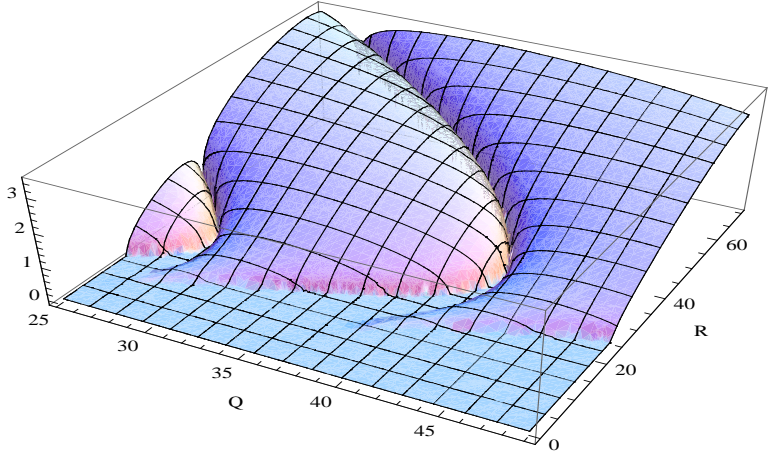}
\end{center}
\caption{The dependence of the real part of characteristic index $\Re \lambda_1$ of the Mathieu equation (\ref{Maa}) upon parameters $Q$ and $R$. This figure is obtained by numerical simulations for different values of parameters, where $Q\in [25,49]$ and $R\in [0,64]$. The grid step  for numerical simulations  equals 0.01 for both parameters.
         }
\label{fig2}
\end{figure}

Depending on the values $q$ and $r$ the solution can grow exponentially or remain bounded, see \cite{UittWatson}.
In Fig. \ref{fig2} the dependence upon the characteristic index $\Re \lambda_1 > 0$ on the parameters $Q$ and $R$ is shown.
\begin{figure}[!t]
\begin{center}
\includegraphics[width=10cm,height=6cm]{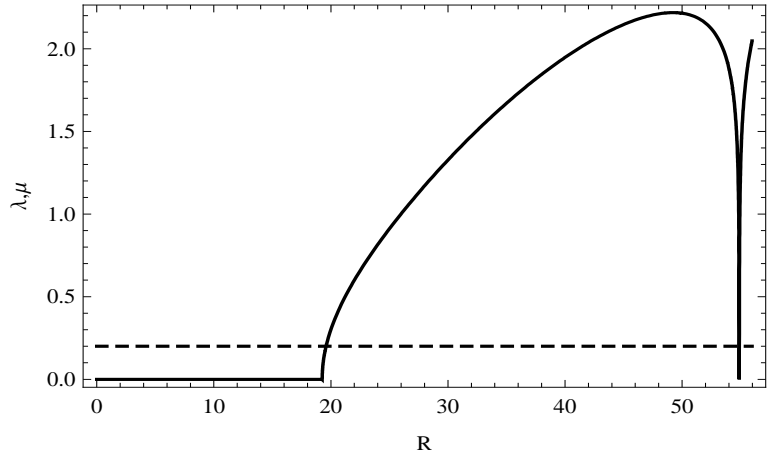}
\end{center}
\caption{A section of the surface of the real part of characteristic index for the Mathieu equation for a fixed $Q=36$. The dashed line is related to the dissipation in equation (\ref{HomMathieu}). The coefficient $\mu$ of dissipation equals 0.1 for numerical simulations.}
\label{fig21}
\end{figure}

A section of the surface $\lambda_1 (Q,R)$ for a fixed value $Q$ and a line corresponding to dissipation are presented in Fig. \ref{fig21}, where
   $Q = 36$,
   $\mu = 0.1$.

In the case under consideration the coefficient $r = r (\tau)$ changes slowly.
In each $2 \pi\,$-interval of the variable $s$ the multiplicator of the solution
of (\ref{HomMathieu}) has the form
$$
   k_{1,2} = \exp (2 \pi (\lambda_{1,2} (q,r (\tau)) - 2 \mu)).
$$
Consider the properties of solutions of (\ref{HomMathieu}) for the fixed value $q = 36$, see Fig. \ref{fig21}.
For $\Re \lambda - 2 \mu <0$ the multiplicators are less than $1$.
Hence it follows that the modulus of the solution decreases on a sequence of intervals.
If $\Re \lambda - 2 \mu > 0$ then the solution growths exponentially on a sequence of intervals.

\begin{figure}[!t]
\begin{center}
\includegraphics[width=10cm,height=6cm]{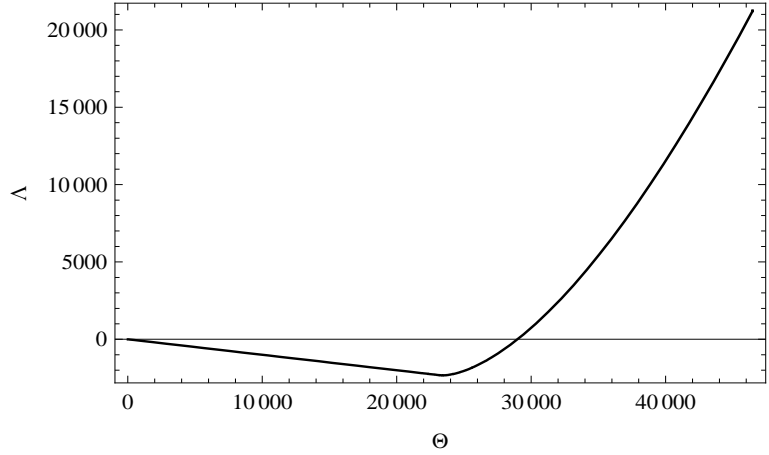}
\end{center}
\caption{The dependence of the integral index $\mathit{\Lambda}$ upon $\theta$.}
\label{fig22}
\end{figure}

In Fig. \ref{fig22} the integral index
$$
   \mathit{\Lambda} (\tau)
 = \int_0^\tau (\Re \lambda_1 (Q, R (\tau')) - 2 \mu)\, d \tau'
$$
is shown depending on time.
The solution of equation (\ref{HomMathieu}) does not depend in the main on the initial data as long as $\mathit{\Lambda} < 0$.
For equation (\ref{nonHomMathieu}) this means that when studying the general solution one can restrict oneself to the particular solution of zero initial data to the inhomogeneous equation and neglect the solution of the homogeneous equation as long as the condition $\mathit{\Lambda} < 0$ is fulfilled.

The correction $y_1$ is determined from the differential equation
$$
   \partial^2_{\theta \theta} y_{1} + y_{1}
 = - 2 \partial^2_{\theta \tau} Y + \delta\, x_0^2.
$$
We look for a solution $y_1$ of the form
$$
   y_1 (\theta,\tau)
 = \ell (\tau) \exp (\imath \theta)
 + \overline{\ell (\tau)} \exp (- \imath \theta),
$$
$\ell$ satisfying the equation
$$
   \partial_\theta \ell
 = \overline{\partial_{\tau} k}\, \exp (- 2 \imath \theta)
 - \partial_{\tau} k
 + \frac{1}{2 \imath}\, \delta x_0^2\, \exp (- \imath \theta).
$$
If $q$ and $r$ are in the domain where the characteristic indices of the Mathieu equation are purely imaginary and
\begin{equation}
\label{antisec}
   \partial_{\tau} k
 = -\,
   \lim_{T \to \infty}
   \frac{\imath \delta}{T}
   \int_0^{T} x_0^2 \exp (- \imath \vartheta) d \vartheta,
\end{equation}
then
$$
   y_1 = o (\theta)
$$
for $\theta \to \infty$.
The averaging over the fast variable on the right-hand side of (\ref{antisec})
determines the derivative of $k$ as function of slow time .
The evolution of $x_0$ over the fast variable is determined by the inhomogeneous
equation (\ref{nonHomMathieu}).
The dependence of $x_0$ upon the slow variable is explained by the slow perturbation of the coefficients of the Mathieu equation.
The explicit dependence of $x_0$ upon the slow variable has not so far been defined.
However, one can verify (\ref{antisec}) by means of numerical simulation of the initial system (\ref{caupOscil}).

The numerical solution of system (\ref{caupOscil}) allows one to evaluate separately the left-hand side and the right-hand side of (\ref{antisec}) and
to determine the relative residual.
We construct a numerical solution for the parameter values $\epsilon = 0.2$,
                                                           $\mathit{\Omega} =1$,
                                                           $\omega = 3$,
                                                           $A = 1$ and
                                                           $\nu = 0.1$.

The system (\ref{caupOscil}) is solved by the Runge-Kutta method of $4\,$th order.
As result we get a numerical solution $x_{\mathrm{num}}$,
                                      $y_{\mathrm{num}}$.
We then divide the entire integration interval into subintervals of length
$2 \pi$ and on each subinterval we compute the Fourier coefficients of
   $\sin \theta$ and
   $\cos \theta$
for the function $\varepsilon y_{\mathrm{num}}$.
From the data obtained in this way we evaluate the difference derivative $k'_\tau$.

The averaging operator on the right-hand side of (\ref{antisec}) is written in the form convenient for analytical computations.
However, in numerical data the dependencies upon fast and slow variables are not
separated from each other.
Hence, it is not possible to directly apply formula (\ref{antisec}).
Instead the averaging operator over the fast variable of (\ref{antisec}) is replaced by the averaging over the interval of length $O (\varepsilon^{-2})$
with centre at the point $\tau_i = \varepsilon^2 t_i$.
In this way we get the values of the right-hand side at the points $\tau_i$.

\begin{figure}[!t]
\begin{center}
\includegraphics[width=8cm,height=6cm]{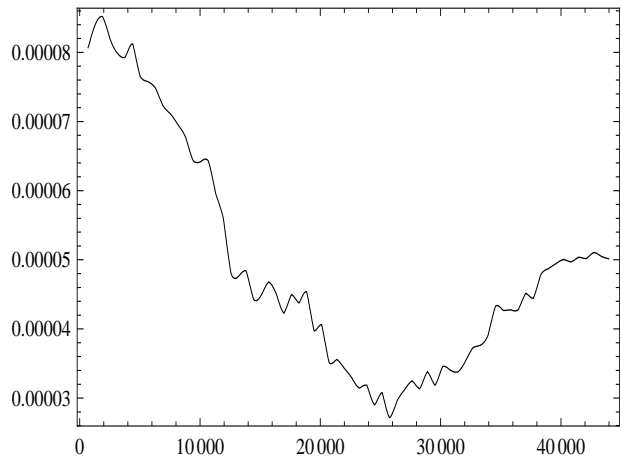}
\end{center}
\caption{The relative error $|S_l-S_r|/|S_l|$ of equation (\ref{antisec}). Here $S_l$ and $S_r$ are
numerical evaluations of the left-hand and right-hand sides of (\ref{antisec}), respectively.}
\label{fig8}
\end{figure}

In Fig. \ref{fig8}
one compares the derivatives in $\tau$ on the left-hand side of equation (\ref{antisec}),
which are evaluated numerically, and the integral on the right-hand side.

%The dashed lines correspond to derivatives.
%
%\begin{figure}[!t]
%\begin{center}
%\includegraphics[width=8cm,height=6cm]{uTauIntbiX1.eps}
%\end{center}
%\caption{The comparison of the real part of the derivative $\Re\, k'_\tau$ with
%that of the integral in (\ref{antisec}).}
%\label{fig6}
%\end{figure}
%
%\begin{figure}[!t]
%\begin{center}
%\includegraphics[width=8cm,height=6cm]{vTauIntbiX1.eps}
%\end{center}
%\caption{The comparison of the imaginary part of the derivative $\Im\, k'_\tau$ with that of the integral in %(\ref{antisec}).}
%\label{fig7}
%\end{figure}
%

We thus conclude that the substitution of the numerical solution for the genuine
solution of (\ref{antisec}) leads to an inessential residual.

\section{Analysis of asymptotic solution for $\lambda \gg 1$}
\label{lambdaBig}
\setcounter{equation}{0}

In this section we carry out the analysis of the behaviour of the main terms of
representation (\ref{secondEx}) under the assumption that
   $\lambda = \omega/\mathit{\Omega} \gg 1$.
This assumption corresponds to the case of forced combinative scattering, where $\lambda$ attains the value $10^2$, see \cite{Khokhlov},
                                        \cite{Akhmanov}.

Equation (\ref{nonHomMathieu}) for the main term $x_0$ can be obviously rewritten in the form
$$
   \frac{1}{\lambda^2}\, \partial^2_{ss} x_0
 + \frac{4 \mu}{\lambda^2}\, \partial_{s} x_0
 + \Big(4 - \frac{2r}{\lambda^2} \cos 2s) x_0
 = \frac{4 f}{\lambda^2}\, \cos \Big( s + \frac{a}{2} \Big).
$$
The asymptotics of the particular solution for large values $\lambda$ is
\begin{eqnarray}
\label{asyMath}
   x_0
 = \frac{2f\, \cos (s + a/2)}{2 - \lambda^{-2} r\, \cos 2s}\, \lambda^{-2}
 + O (\lambda^{-4}).
\end{eqnarray}

Our next objective is to treat equation (\ref{antisec}) for $\lambda \gg 1$.
To this end we substitute (\ref{asyMath}) into equation (\ref{antisec}) and evaluate the integral explicitly.
As result we get
\begin{eqnarray}
\label{anti}
\frac{d K_1}{d \tau'} & = & \,\,\,\,\, \frac{\partial H}{\partial K_2},
\nonumber
\\
\frac{d K_2}{d \tau'} & = &          - \frac{\partial H}{\partial K_1},
\nonumber
\\
\end{eqnarray}
where $\tau' = \lambda^{-8} \tau$.
This is a Hamiltonian system with Hamiltonian
$$
   H
 = \frac{A^2}{2 \pi \mathit{\Omega}^3}\,
   \frac{|K|^2 - K_1}{K_1 (1 - |K|^2) - (|K|^2 - K_1) \sqrt{1 - |K|^2}}.
$$

\begin{figure}[!t]
\begin{center}
\includegraphics[width=8cm,height=6cm]{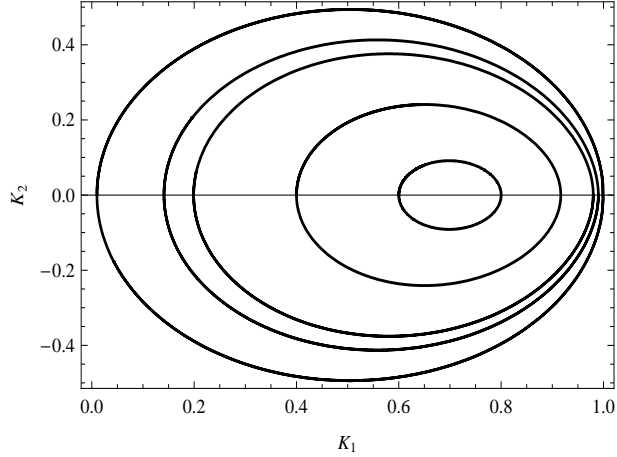}
\end{center}
\caption{A phase portrait of solutions of (\ref{anti}) in a neighborhood of `centre' $K_1=1/\sqrt{2}, K_2=0$. This figure is obtained numerically with diverse initial data for system (\ref{anti}).}
\label{fig10}
\end{figure}

The system (\ref{anti}) is easily seen to have a stable equilibrium point of type `centre'
$$
\begin{array}{rcl}
   K_1 & = & \displaystyle \frac{1}{\sqrt{2}},
\\
   K_2 & = & 0,
\end{array}
$$
see Fig. \ref{fig10}.

\begin{figure}[!t]
\begin{center}
\includegraphics[width=8cm,height=6cm]{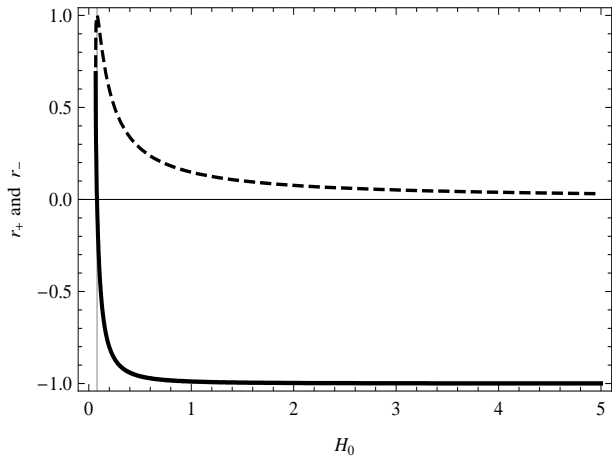}
\end{center}
\caption{Dependence  of $r_+$ (dashed) and $r_-$ (solid) evaluated by (\ref{rpm}) on $H_0$. The vertical line corresponds to the value $H_0=1/(4\pi)$. An intersection of these lines allows one to estimate the maximal value of envelope amplitude.}
\label{fig11}
\end{figure}

The amplitude of oscillations can be found from the system
\begin{eqnarray}
   H (K_1,K_2)
 & = &
   H_0, \label{large-small}
\\
   \frac{d}{d \tau'}\, (K_1^2 + K_2^2)
 & = &
   0.\nonumber
\end{eqnarray}
In Fig. \ref{fig11} the dependence of the largest $r_+$ (dashed curve) and
                                          smallest $r_-$ (solid curve)
values of solutions of (\ref{large-small}) upon the Hamiltonian is shown. The vertical line corresponds to $H_0=1/(4\pi).$ 
This allows one to evaluate the maximal value of the amplitude of oscillation envelope through
$$
   \frac{\lambda^2}{\varepsilon}\, \max \{ |r_-|, |r_+| \},
$$
where
\begin{equation}
   r_\pm
 = \frac{(A^2 - 2 \pi \mathit{\Omega}^3 H_0)
     \pm \sqrt{- A^4 + 4 A^2 \pi \mathit{\Omega}^3 H_0
                     + 4 \pi^2 \mathit{\Omega}^6 H_0^2}}
        {4 \pi \mathit{\Omega}^3 H_0}. \label{rpm}
\end{equation}
The period is determined by
\begin{equation}
   T (H_0)
 \sim \frac{\lambda^8}{\varepsilon^{2}}\,
      \int_{c} \frac{d K_{1}}{\displaystyle \frac{\partial H}{\partial K_{2}}}, \label{period}
\end{equation}
where the integration is over the cycle $c$ in the complex plane of the variable
$K = K_1 + \imath K_2$ given by
$$
   \frac{|K|^2  -  K_1}
        {K_1 (1 - |K|^2) - (|K|^2 - K_1) \sqrt{1-|K|^2}}
 = - \frac{2 \pi \mathit{\Omega}^3 H_0}{A^2}.
$$

\begin{figure}[!t]
\begin{center}
\includegraphics[width=8cm,height=6cm]{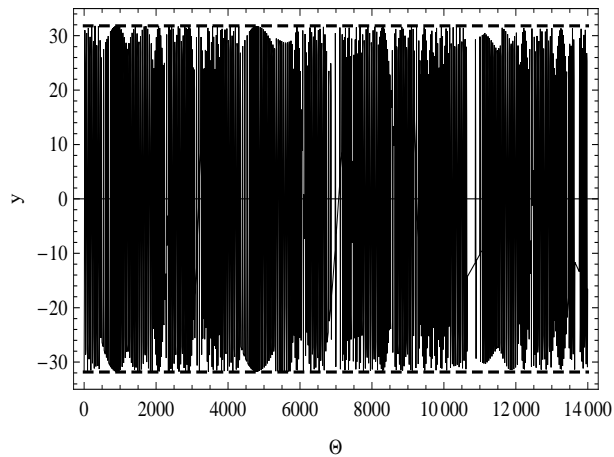}
\end{center}
\caption{The constant envelope function and oscillations of the component $y$. 
The component $y$ is obtained by numerical simulations for the solution of system (\ref{caupOscil}). The constant envelope function (dashed curve) equals  $\pm \lambda^2 \varepsilon^{-1}/\sqrt{2}$ which is  related to the  stationary solution $K_1=1/\sqrt{2}, K_2=0$ of Hamiltonian system (\ref{anti}).}
\label{fig12}
\end{figure}
In Fig. \ref{fig12} one sees the results of numerical simulation for the solution of system (\ref{caupOscil}) with initial data corresponding to the
stationary solution
$$
\begin{array}{rclccrcl}
   x (0)
 & =
 & \displaystyle \frac{f}{\lambda^2},
 &
 &
 & y (0)
 & =
 & \displaystyle - \frac{\lambda^2}{\varepsilon \sqrt{2}},
\\
   x' (0)
 & = &
   0,
 &
 &
 & y' (0)
 & =
 & 0.
\end{array}
$$

The analytical result obtained in the paper agrees well to the numerical simulation.
For the values of parameters $K_1$ and $K_2$, which are used in the numerical simulation, the Hamiltonian value proves to be
   $H_0 = 1/ (4 \pi)$.
In this case the amplitude of pulsations is evaluated by    $\lambda^2 \varepsilon^{-1}$. It is shown in Fig. \ref{fig9}. The horizontal line corresponds to  $\lambda^2 \varepsilon^{-1}$.

The explicit value for the period of envelope oscillations can be evaluated from (\ref{period}). Our numerical simulations with  parameters $A=1, \omega=3, \Omega=1, \mu=0.1, \varepsilon=0.2$ give  the value $T=5264.76$ in the variable $\tau$.

\begin{figure}[!t]
\begin{center}
\includegraphics[width=8cm,height=6cm]{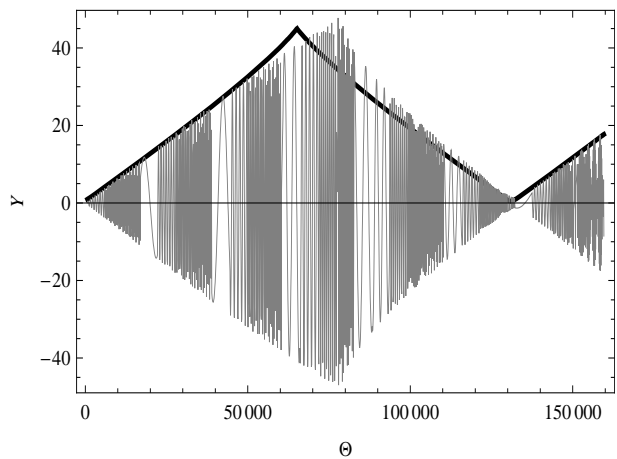}
\end{center}
\caption{The numerically evaluated $y\,$-component of the solution and the envelope. The envelope function is evaluated as the solution of (\ref{anti}).}
\label{fig13}
\end{figure}

\begin{figure}[!t]
\begin{center}
\includegraphics[width=8cm,height=6cm]{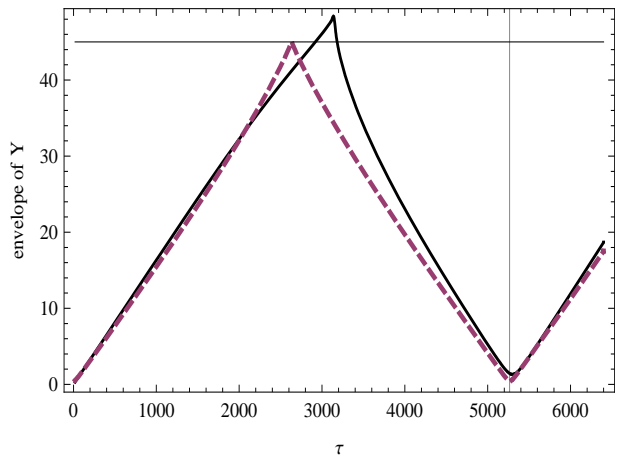}
\end{center}
\caption{Numerically evaluated envelope function of $y\,$-component of the solution (solid curve) and the approximation of the solution envelope as solution of system (\ref{anti}) (dashed curve). The vertical line corresponds to the value of the period $T=5264.76$. It is determined numerically by (\ref{period}). The horizontal line is an estimate of maximal value $\lambda^2\varepsilon^{-1}$ of envelope amplitude.}
\label{fig9}
\end{figure}

In Fig. \ref{fig13} the pulsations of the $y\,$-component are shown as well as
the envelope superposed on them.
The envelope is computed numerically as solution of system (\ref{anti}) with
   $\varepsilon = 0.2$,
   $\lambda = 3$,
   $A =1$.
All these numerical evaluations are realized by the Adams method with the precision $10^{-10}.$

System (\ref{anti}) gives an approximation for determining the behaviour of functions $K_1$ and $K_2$.
In Fig. \ref{fig9} the solid curve corresponds to the envelope of the $y\,$-component of the solution.
This curve is found from a numerical  solution of system
(\ref{caupOscil}).
The dashed curve $\sqrt{K_1^2+K_2^2}$ gives an approximation of the solution envelope as solution of system (\ref{anti}). A point $K_1=0, K_2=0$ is an unstable node equilibrium point. Initial data for system (\ref{anti}) correspond to the linear resonance of $y\,$-component of the solution for (\ref{caupOscil}) in the initial stage, see Fig. \ref{fig1a}. In Fig. \ref{fig9}  the vertical line shows the value of the period $T=5264.76$. It is determined numerically by expression (\ref{period}). The horizontal line is an estimate of the maximal value $\lambda^2\varepsilon^{-1}$ of envelope amplitude.

\section{Conclusion}
\label{s.conclusion}
\setcounter{equation}{0}
It is shown in the paper that for the large time $t \sim \varepsilon^{-2}$ with
$\varepsilon$ being a coupling parameter the behaviour in the main of a resonantly perturbed system of coupled oscillators is described by a Hamiltonian
system. 
This system is studied in the physically interesting case when the quotient of oscillator frequencies $\omega / \mathit{\Omega}$ is large enough.
Certain pulsations prove to occur in the solution of the initial system of coupled oscillators.
The Hamiltonian system obtained allows one to evaluate the period and amplitude
of these pulsations.

\bigskip

{\bf Acknowledgements\,}
The research was supported by
   the RFBR grant 09-01-92436-KE-a,
   the DFG grant TA 289/4-1
and by
   grant 2215.2008.1 for Russian scientific schools.
The first author wishes to thank the DAAD and Ministry of Education and Science of the RF for financial support in the framework of program ``Mikhail Lomonosov.''

\end{document}